\documentclass[prl, twocolumn, superscriptaddress]{revtex4-1}
\usepackage{graphicx}
\usepackage[]{lineno}
\usepackage{setspace}
\usepackage{amsmath}
\usepackage{xcolor}
\usepackage{float}
\usepackage[colorlinks,
            urlcolor=blue,
            linkcolor=blue,
            anchorcolor=blue,
            citecolor=blue]{hyperref}
\renewcommand{\section}[1]{{\par\it #1.---}\ignorespaces}
\let\oldequation\equation
\let\oldendequation\endequation

\renewenvironment{equation}
{\linenomathNonumbers\oldequation}
{\oldendequation\endlinenomath}
\begin{document}
\title{Arbitrarily configurable nonlinear topological modes}
\author{Kai Bai}
\affiliation{Key Laboratory of Artificial Micro- and Nano-structures of Ministry of Education and School of Physics and Technology, Wuhan University, Wuhan 430072, China}
\author{Jia-Zheng Li}
\affiliation{Key Laboratory of Artificial Micro- and Nano-structures of Ministry of Education and School of Physics and Technology, Wuhan University, Wuhan 430072, China}
\author{Tian-Rui Liu}
\affiliation{Key Laboratory of Artificial Micro- and Nano-structures of Ministry of Education and School of Physics and Technology, Wuhan University, Wuhan 430072, China}
\author{Liang Fang}
\affiliation{Key Laboratory of Artificial Micro- and Nano-structures of Ministry of Education and School of Physics and Technology, Wuhan University, Wuhan 430072, China}
\author{Duanduan Wan}
\email{ddwan@whu.edu.cn}
\affiliation{Key Laboratory of Artificial Micro- and Nano-structures of Ministry of Education and School of Physics and Technology, Wuhan University, Wuhan 430072, China}
\author{Meng Xiao}
\email{phmxiao@whu.edu.cn}
\affiliation{Key Laboratory of Artificial Micro- and Nano-structures of Ministry of Education and School of Physics and Technology, Wuhan University, Wuhan 430072, China}
\affiliation{Wuhan Institute of Quantum Technology, Wuhan 430206, China}

\begin{abstract}
Topological modes (TMs) are typically localized at boundaries, interfaces and dislocations, and exponentially decay into the bulk of a large enough lattice. Recently, the non-Hermitian skin effect has been leveraged to delocalize the wavefunctions of TMs from the boundary and thus to increase the capacity of TMs dramatically. Here, we explore the capability of nonlinearity in designing and reconfiguring the wavefunctions of TMs. With growing intensity, wavefunctions of these in-gap nonlinear TMs undergo an initial deviation from exponential decay, gradually merge into arbitrarily designable plateaus, then encompass the entire nonlinear domain, and eventually concentrate at the nonlinear boundary. Intriguingly, such extended nonlinear TMs are still robust against defects and disorders, and stable in dynamics under external excitation. Advancing the conceptual understanding of the nonlinear TMs, our results open new avenues for increasing the capacity of TMs and developing compact and reconfigurable topological devices. 
\end{abstract}

\maketitle
\section{Introduction}\label{introduction}
The concept of topological matters flourished rapidly in various fields such as condensed matters \cite{1Wang2017}, photonics \cite{2Wang2009,3Rechtsman2013,4Lu2014,5PhysRevLett.100.013904,6Hafezi2013,7Khanikaev2013,8Mittal2016,9PhysRevX.5.031011}, circuits \cite{10PhysRevX.5.021031,11PhysRevLett.114.173902,12Imhof2018,13Lee2018}, and acoustic and mechanical systems \cite{14Xiao2015a,16Ma2019,17Huber2016,18Wang2022,19Hu2021,20PhysRevLett.113.175503}. As ensured by the conventional bulk-boundary correspondence \cite{21PhysRevLett.71.3697}, topologically nontrivial bulks give rise to topological modes (TMs) localized at the boundary, interface, and crystallographic defects such as dislocations and disclinations \cite{22Lin2023}. These TMs are robust against disorders and backscattering immunity to certain defects. Numerous novel phenomena and potential applications rooted in TMs have been elucidated in the past decades \cite{23Hafezi2011,24Hu2021,25Harari2018a,26Bandres2018,27Zeng2020,28Barik2018,29Blanco-Redondo2018,30Mittal2018}. However, as inherited from the bulk-boundary correspondence, TMs exponentially decay into the bulk and hence have limited capacity. The requirement of bulky topological materials and the limited available capacity of TMs exhibit a bottleneck in potential applications. Recently, the non-Hermitian skin effect \cite{31PhysRevLett.121.086803,32PhysRevX.8.041031,33PhysRevLett.121.026808} has been leveraged to delocalize TMs \cite{18Wang2022,34PhysRevB.103.195414}. Therein, the nonreciprocal coupling tunes the TMs into completely extended modes. Here, we explore the miraculous consequence of nonlinearity in reconfigurating TMs. Our work demonstrates the capability of harnessing nonlinearity to reshape TMs into arbitrarily designed profiles such as square, isosceles triangular, and sinusoidal waves. In addition, instead of a fixed mode profile for each sample in previous works \cite{18Wang2022,34PhysRevB.103.195414}, the lattices covered by the extended TMs herein can be easily tuned with the input intensity. 

Nonlinearity is ubiquitous \cite{35RevModPhys.88.035002,36Smirnova2020b,37Bai2023,38PhysRevLett.130.266901,39Assawaworrarit2017,40PhysRevLett.127.076802}, which, coupled with topology, can lead to exciting physics and novel phenomena \cite{41PhysRevLett.117.143901,42PhysRevX.11.041057,43PhysRevLett.111.243905,44Mukherjee2020}. The topological edge \cite{41PhysRevLett.117.143901,42PhysRevX.11.041057} and bulk solitons \cite{43PhysRevLett.111.243905,44Mukherjee2020} were discovered in nonlinear topological insulators. They are strongly self-localized and propagate unidirectionally along the edge or inside the bulk when the nonlinear effects compensate for the dispersion. Thanks to the inherent reconfigurability of nonlinear structures, the excitation intensity can induce topological phase transitions, leading to the emergence of topologically robust edge states \cite{45PhysRevB.93.155112,46Hadad2018,47Maczewsky2020} and corner states\cite{48Kirsch2021,49PhysRevLett.123.053902}. Recently, nonlinear effects have been extended to non-Hermitian topological insulators for active tuning of parity-time symmetry and the corresponding topological edge states \cite{50Xia2021a}. These above works pave the way for reconfigurable devices imbued with topological features. Nevertheless, the question of whether nonlinearities may design and reconfigure TMs has remained largely uncharted. Clearly, it would be of great interest to pursue amoeboid nonlinear TMs which are robust against disorders as protected by a nontrivial topology while uniquely controllable through external sources as inherited from the reconfigurability of nonlinearity. Exploring nonlinear topological physics continues to be an intriguing frontier yet to be fully unveiled.

Here, we leverage nonlinearity to deform, reshape, and design the wavefunctions of TMs. Our system consists of a one-dimensional lattice that is linear and topologically nontrivial and a nonlinear one that features alternating linear and nonlinear couplings \cite{45PhysRevB.93.155112,46Hadad2018,47Maczewsky2020,48Kirsch2021}. In the low-intensity regime (where the nonlinearity is negligible), the nonlinear chain remains topologically trivial, supporting a topological zero mode (TZM) localized at the interface. With increasing intensity, the profile of the TZM is deformed on the nonlinear lattice and deviates from the exponential decaying behavior. As the intensity is above a certain threshold, the TZM merges into an arbitrarily designable plateau that gradually covers the entire nonlinear lattice domain. Interestingly, the eigenfrequencies of TZMs stay at the gap center of the nonlinear eigenfrequency spectrum. The topology, which guarantees the existence of a TZM of the perturbed system, is characterized by the nonlinear spectral localizer \cite{51PhysRevB.108.195142,52Cheng2023,53PhysRevLett.131.213801}. When excited, TZMs are stable in dynamics against noise while tunable with the excitation power. Our findings promote the understanding of reconfigurable TZMs and open new avenues for utilizing the reconfigurability in quantum and classical nonlinear systems.
\begin{figure*}
	\centering
	\includegraphics[width=1.5\columnwidth]{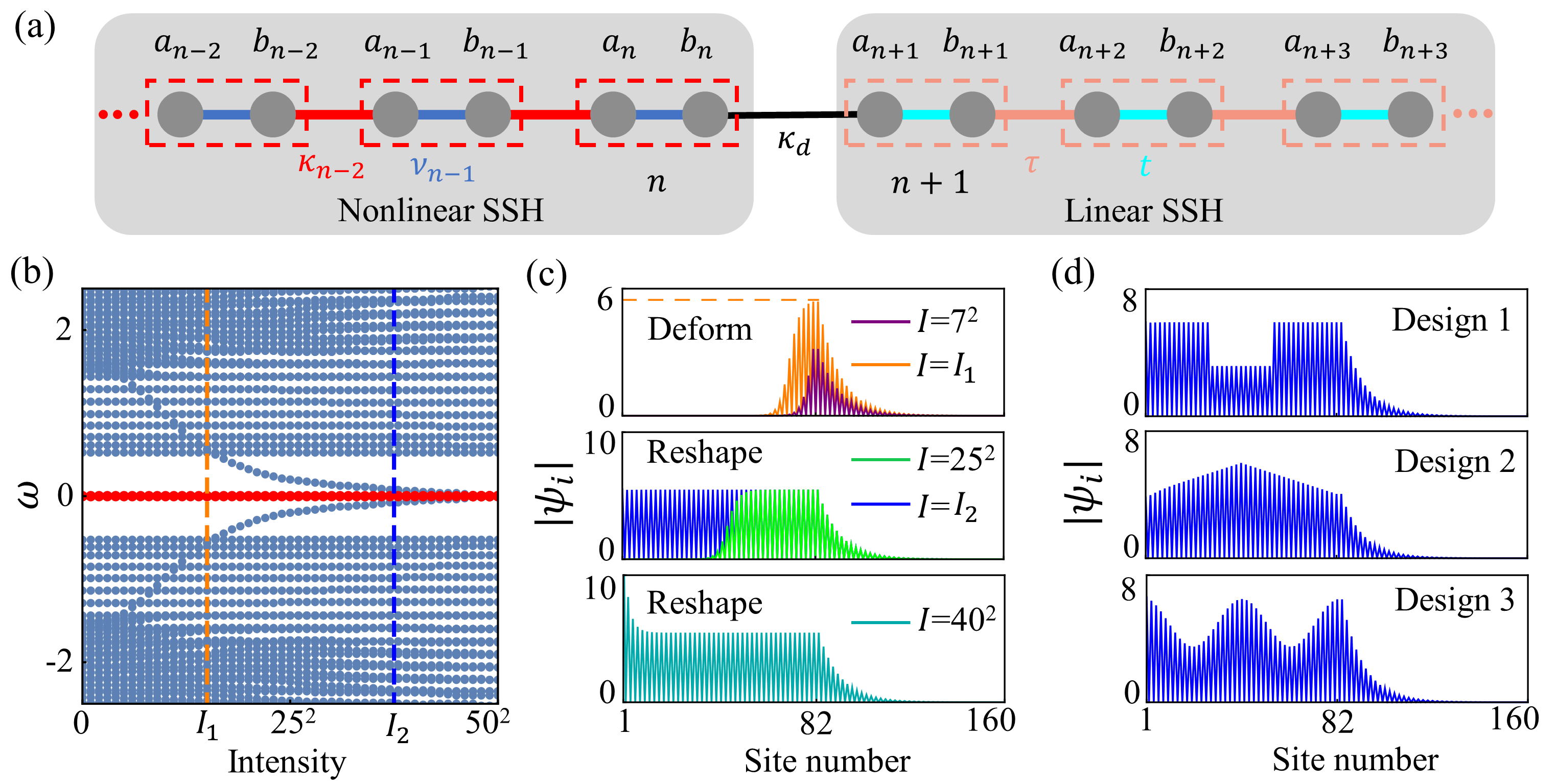}\\
	\caption {(a) Schematic of a tight-binding lattice consisting of a nonlinear (left) and linear (right) SSH chain. $a_i$ and $b_i$ are the field amplitudes at different sites of the $i-th$ unit cell (marked by the dashed boxes). The red bonds denote nonlinear intercell couplings $\{\kappa_{i-1} (b_{i-1},a_i)\}$ that depend on the intensities of sites connected by the bonds. All the other couplings are linear with the hoppings being $\{\nu_i\}$, $\kappa_d$, $t$ and $\tau$ for the blue, black, cyan and pink bonds, respectively. (b) Evolution of the nonlinear eigenfrequency spectrum of a finite lattice as the intensity $I$ is increased. The red dots mark the TZM, whose eigenfrequency is pinned at zero independent of $I$. (c) The wavefunctions of the TZM for different $I$. At $I_1=15^2$, the wavefunction emerges a plateau. The orange horizontal dashed line is for eye guiding. This plateau will gradually extend to the entire nonlinear domain at $I_2=37.5^2$ and eventually the field concentrates at the nonlinear boundary. (d) Based on Eq. \eqref{eq6}, the shape of the plateau (at the corresponding $I_2$) can be turned into an arbitrary shape, such as a square, isosceles triangle, cosine, etc. The parameters used are $t=2$, $\tau=\kappa_d=2.5$, $\left\{\nu_i\right\}=2.5$, $\kappa_i={\widetilde{\kappa}}_i+\alpha (\left|a_i\right|^2+\left|b_{i-1}\right|^2)$ with $\alpha=0.05$. $n=41$ and the total number of lattice sites is 161. In (b) and (c), ${{\widetilde{\kappa}}_i}=1$. In (d), the corresponding distribution of ${{\widetilde{\kappa}}_i}$ are provided in the Supplemental Material, Sec. 2.}
	\label{fig1}
\end{figure*}

\section{Shapeshifting of TZMs with nonlinearity}
Our system consists of a nonlinear Su–Schrieffer–Heeger (NL-SSH) chain \cite{45PhysRevB.93.155112} and a linear one \cite{54PhysRevLett.42.1698} as sketched in Fig. \ref{fig1}(a). The nonlinear Schrödinger equation in real space is 
\begin{equation}
H_{\left|\boldsymbol{\psi}\right\rangle}\left|\boldsymbol{\psi}\right\rangle=\omega\left|\boldsymbol{\psi}\right\rangle, \label{eq1}
\end{equation} where $\omega$ is the eigenfrequency. $\left|\psi\right\rangle\equiv\left(\cdots,a_i,b_i,\cdots\right)^T$ is the eigenstate with superscript $T$ short for transpose, and $a_i$, $b_i$ representing the field amplitudes at different sublattices of the $i-th$ unit cell. The tight-binding Hamiltonian $H_{\left|\boldsymbol{\psi}\right\rangle}$ is
\begin{eqnarray}
H_{\left|\boldsymbol{\psi}\right\rangle}=&&\sum_{i<n}(\nu_{i}|a_{i}\rangle\langle b_{i}|+\kappa_{i-1}|a_{i}\rangle\langle b_{i-1}|)+\kappa_{d}|a_{n+1}\rangle\langle b_{n}|\nonumber\\&&+\sum_{i>n}(t|a_{i}\rangle\langle b_{i}|+\tau|a_{i+1}\rangle\langle b_{i}|)+H.c., \label{eq.2}
\end{eqnarray}
The nonlinear chain has $n$ unit cells, and $\nu_i$ and $\kappa_i$ represent the corresponding intracell and intercell couplings, respectively. $\kappa_i={\widetilde{\kappa}}_i+\alpha(\left|a_{i+1}\right|^2+\left|b_i\right|^2)$ with a linear term ${\widetilde{\kappa}}_i$ and a Kerr nonlinear coefficient $\alpha$ \cite{45PhysRevB.93.155112,46Hadad2018,47Maczewsky2020,48Kirsch2021,49PhysRevLett.123.053902}. $t$ and $\tau$ denote the intracell and intercell coupling in the linear chain, respectively. $\kappa_d$ is the coupling at the interface. Given the specific form of nonlinearity, the eigenvalues and eigenstates of the nonlinear Schr\"{o}dinger equation can be solved numerically using a self-consistent method \cite{49PhysRevLett.123.053902}. Figure  \ref{fig1}(b) shows the eigenvalue distribution versus the total wavefunction intensity $I=\langle\psi\left|\psi\right\rangle=\sum_{i}(\left|a_i\right|^2+\left|b_i\right|^2)$. The red dots mark the states with zero eigenfrequencies inside the gap. There are another two modes (dark blue dots) that gradually merge into the gap and approach the zero frequency. These two modes originate from the bulk modes inside the nonlinear lattice region, the evolution of the nonlinear part of these wavefunctions is similar to the topological edge states introduced through nonlinearity-induced topological transitions \cite{45PhysRevB.93.155112,46Hadad2018} (See detailed discussion in the Supplemental Material, Sec. 1) 

When $I$ is small (the nonlinear effects are negligible), the NL-SSH remains topologically trivial $(\nu_i>{\widetilde{\kappa}}_i)$, and forms an interface with the topologically nontrivial linear SSH $(\tau>t)$. Therefore, there is a TZM localized at the interface and exponentially decaying toward the chains on both sides. With the increasing of $I$, the wavefunction of the TZM deviates from exponential decay on the nonlinear lattice [see Fig. \ref{fig1}(c), at $I=7^2$ for example]. Subsequently, the wavefunction exhibits a plateau at $I=I_1$ [the orange solid line in Fig. \ref{fig1}(c), and the orange dashed line indicates the plateau]. This plateau then gradually extends to cover the whole nonlinear lattice domain when $I>I_1$ [see Fig. \ref{fig1}(c), the green line at $I={25}^2$ for example], and eventually fills the entire NL-SSH chain at $I=I_2$ [the blue line Fig. \ref{fig1}(c)]. When $I$ is further increased to be above $I_2$, the wavefunction will concentrate at the left boundary of the nonlinear lattice while the plateau built before is preserved [see Fig. \ref{fig1}(c), the cyan line at $I={40}^2$ for example]. Interestingly, the plateau can be designed arbitrarily. Similar to Fig. \ref{fig1}(c), the wavefunction of the TZM starts with an interface state and gradually extends to cover the whole nonlinear lattice under the increasing of $I$. Figure \ref{fig1}(d) shows three different profiles, square, isosceles triangle, and cosine, of the TZMs at their corresponding critical intensity $I=I_2$. Furthermore, the magnitude at the linear part of TZMs’ wavefunction is tunable with $\kappa_d$. The details are provided in the Supplemental Material, Sec. 2.  

\section{Topological protection of the TZMs}
\begin{figure}
	\centering
	\includegraphics[width=1.0\columnwidth]{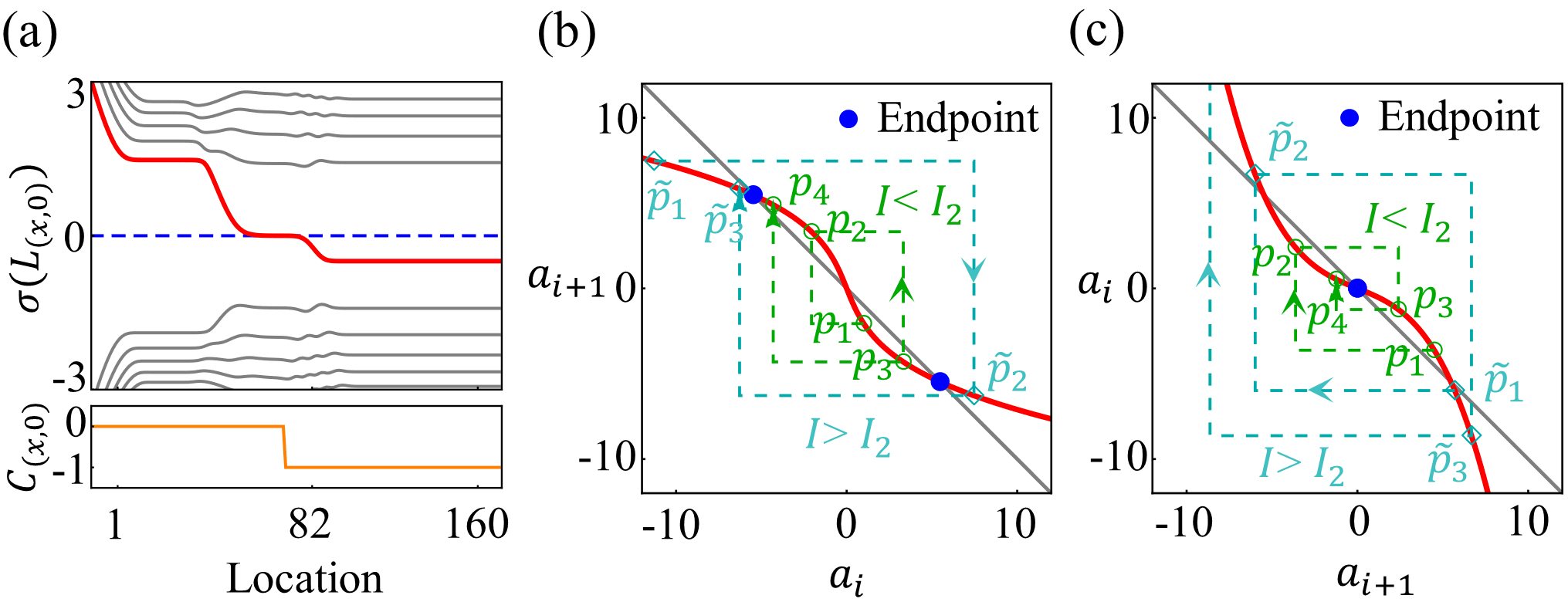}\\
	\caption {(a) Eigenvalues of the spectral localizer $\sigma(L_\lambda)$ and the topological invariant $C_\lambda$ versus the location $x$, where the solid red line corresponds to the smallest singular value, i.e., the eigenvalue closest to zero. (b, c) The schematic of iterative trajectories of $\{a_i\}$ as the site number increases (b) or decreases (c). The red lines are given by Eq. \eqref{eq6} and the gray lines correspond to $a_i=-a_{i+1}$. The dark green and cyan dashed lines indicate two iteration trajectories corresponding to $I<I_2$ and $I>I_2$. In (a), $I={25}^2$, $\beta=0.2$, and the lattice constant and energy (frequency) unit are set as 1. Other parameters are the same as those in Fig. \ref{fig1}(c). }
	\label{fig2}
\end{figure}

Generally, nonlinear effects are intrinsically local. When the nonlinear effects are strong, the spatial periodicity of $H_{\left|\psi\right\rangle}$ will be broken, and the conventional topological invariant defined in the momentum space becomes ill-defined. As such, active research is ongoing to understand the nonlinear topological property. Here, we adopt a nonlinear spectral localizer \cite{51PhysRevB.108.195142,52Cheng2023} to characterize the topological origin of the TZMs. The spectral localizer is a Hermitian composite operator $L_\lambda$ that combines $H_{\left|\psi\right\rangle}$ with the information of the real-space position operators $X$ using a non-trivial Clifford representation,
\begin{equation}
L_{\lambda\equiv(x,\widetilde{\omega})}(X,H_{\left|\psi\right\rangle})=\beta(X-x)\otimes\Gamma_{x}+(H_{\left|\psi\right\rangle}-\widetilde{\omega})\otimes \Gamma_{y}.\label{eq.3}
\end{equation}
Here, $\Gamma_{x,y}$ is Pauli matrice, $\beta$ is a hyperparameter to ensure that the units are comparable, $X\equiv{x_i}$ where $x_i$ denotes the coordinate of the $i-th$ lattice site. At a specified location $x$ and frequency $\widetilde{\omega}$ (marked as $\lambda\equiv(x,\widetilde{\omega})$, which can be any value, even outside of the system’s spatial and spectral regions), the local topological invariant is given by
\begin{equation}
C_\lambda=\frac{1}{2}sig(L_\lambda),\label{eq.4}
\end{equation}
where $sig$ is the signature of a matrix, i.e., its number of the positive eigenvalues minus that of the negative ones. Figure \ref{fig2}(a) shows the eigenvalues of $L_\lambda$ [denoted as  $\sigma(L_\lambda)$] and the corresponding $C_\lambda$ at $\widetilde{\omega}=0$. $C_\lambda$ changes when one of $\sigma(L_\lambda)$ crosses zero such as the red line in the upper panel of Fig.  \ref{fig2}(a). When the red line crosses zero at $x_0$ [$det\left(L_\lambda\right)=0$], the system exhibits a state approximately localized at $\lambda=\left(x_0,0\right)$, thus realizing the bulk-boundary correspondence, i.e., the change of $C_\lambda$ corresponds to a TZM.
The smallest singular value $\mu_\lambda=min[|\left.\sigma(L_\lambda\right)|]$ of the spectral localizer provides additional information of the system at $\lambda$. Small values of $\mu_\lambda$ allow the existence of a state approximately localized near $\lambda$, while large ones indicate that the system does not support such a state. Therefore, $\mu_\lambda$ can be thought of as a “local band gap”, and the topological protection of the TZMs can be characterized by 
\begin{equation}
||\Delta H(\delta)||\leq \mu_{\lambda}^{max},\label{eq5}
\end{equation}
where $\Delta H(\delta)$ is the largest singular value of $\Delta H(\delta)=H_{|\boldsymbol{\psi}\rangle}(\delta)-H_{|\boldsymbol{\psi}\rangle}$, with $H_{|\boldsymbol{\psi}\rangle} (\delta)$ representing the perturbed nonlinear Hamiltonian. Let $\mu_{\lambda}^{max}\equiv\underset{{\bf x}}{\text{max}}[\mu_{(x,0)}(X,H_{\left|\boldsymbol{\psi}\right\rangle})]$ denotes the maximum $\mu_{\left(x,0\right)}$ inside the topological domain ($C_\lambda\neq0$). As long as Eq. \eqref{eq5} is satisfied, the topological protection guarantees the existence of a nonlinear eigenmode with an eigenvalue of 0, i.e., a TZM. 

The eigenstates of the TZM in the nonlinear part satisfy
\begin{equation}
\nu_ia_i+\left({\widetilde{\kappa}}_i+\alpha a_{i+1}^2\right)a_{i+1}=0,\label{eq6}
\end{equation}
and all $b_i=0$ due to the bipartite property. Figure \ref{fig2}(b) sketches the iterative trajectories of $\{a_i\}$ for two different paths as the site number increases. Regardless of whether the starting point has a smaller amplitude (such as $p_1$, corresponding to $I<I_2$) or a larger amplitude (such as $\widetilde{p}_1$, corresponding to $I>I_2$), it will eventually converge to endpoints with a non-zero amplitude value, marked by the blue dots. Here, the two blue points correspond to the plateau in Fig. \ref{fig1}(c), and the corresponding height of the platform is
\begin{equation}
\left|a_i\right|=\sqrt{(\nu_i-{\widetilde{\kappa}}_i)/\alpha}.\label{eq7}
\end{equation}
The intensity $I$ as well as the coefficients $\{\nu_i\}$, ${{\widetilde{\kappa}}_i}$ and $\alpha$ affect how fast the iteration converges and thus the width of the plateau. Figure \ref{fig2}(c) shows the iterative trajectories of $\{a_i\}$ along the direction of decreasing the site number. Amplitudes starting at less than the plateau (such as $p_1$, corresponding to $I<I_2$) converge to the endpoint with vanishing amplitude, which corresponds to the decaying of $\{a_i\}$ towards the left boundary of the NL-SSH chain. While amplitudes starting at larger than the plateau (such as ${\widetilde{p}}_1$, corresponding to $I>I_2$) will diverge at the left boundary, and the $I={40}^2$ curve in Fig. \ref{fig1}(c) is one such case. The waveform of TZM in the linear chain can be handled similarly, and the magnitude is tunable with $\kappa_d$ through $a_n\nu_n+a_{n+1}\kappa_d=0$. (See details in the Supplemental Material, Sec. 2.) Furthermore, when we vary ${{\widetilde{\kappa}}_i}$, the shape of the plateau can be designed arbitrarily based on Eq. \eqref{eq6} [see Fig. \ref{fig1}(d) and the Supplemental Material, Sec. 3 for a systematic approach.]

\section{Stability of the TZMs under external excitation}
\begin{figure}
	\centering
	\includegraphics[width=1.0\columnwidth]{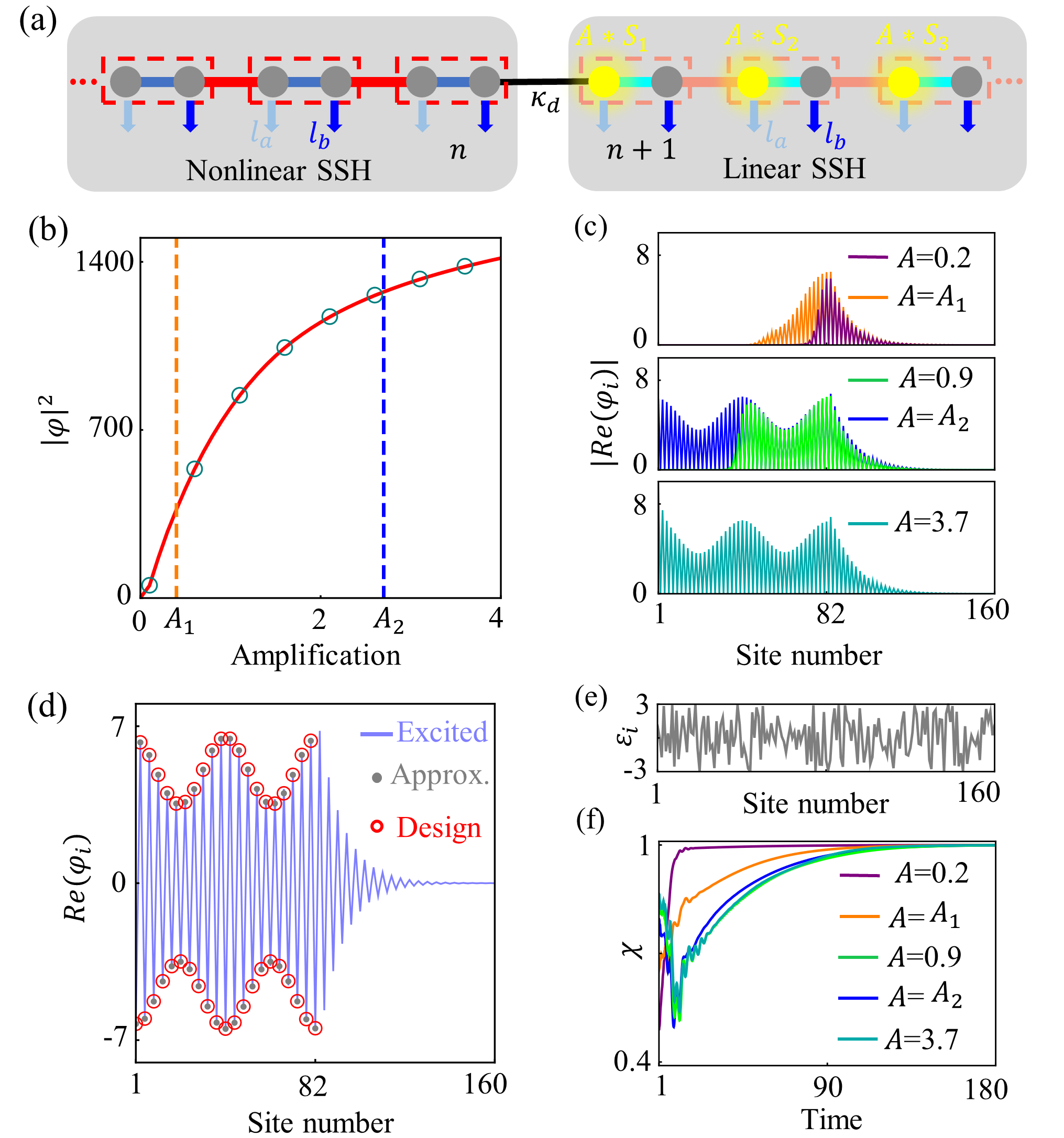}\\
	\caption {(a) Schematic of excitation. These yellow spots represent the locations of the sources, all of which are located in the linear chain, and the number of excitation sources can be reduced to one. The corresponding distribution is $\{A\ast S_i\}$ where $A$ denotes a global amplification factor. $l_a$ and $l_b$ are the losses at different sites, which is unavoidable in real systems and can be used to stabilize the excited state. (b) The intensity $\left|\varphi\right|^2$ of the excited state versus $A$, where $A_1=0.4$ and $A_2=2.7$. The dark cyan circles obtained from Eq. \eqref{eq8} coincide with the solid red line solved using the self-consistent Green’s function [Eq. (S7)]. (c) The wavefunctions at different $A$. (d) The excited wavefunction (light blue line) is almost perfectly consistent with the results from Eq. \eqref{eq9} (gray dots, the approximate wavefunction) and Eq. \eqref{eq6} (red circles, the wavefunction of targeted TZM). (e) A typical distribution of noise. (f) Evolution of the similarity function $\chi$ for different $A$. In the presence of noise, $\chi$ deviates from 1 and will fall back to 1 eventually. Thus, we can obtain stable excited states almost identical to TZMs. Here, $\widetilde{\omega}=0$, $l_a=0.01$ and $l_b=0.5$, and the distribution of $\{S_i\}$ is provided in Fig. S4(b). Other parameters are the same as those in Design 3 of Fig. \ref{fig1}(d).}
	\label{fig3}
\end{figure}
Compared with conventional linear topological structures, nonlinear ones exhibit inherent reconfigurability. In particular, the dynamics of TZMs depend on the intensity, which offers a unique controllability utilizing external sources. However, it still remains a fundamental challenge to reach the designed TZMs in practical use. Firstly, the nonlinear TZM depends delicately on the local field distributions. Secondly, the excited state in general is a composition of different modes of the corresponding Green’s function, and such a composition typically deviates from the TZM. Thirdly, one needs to exert additional effort to stabilize the excited mode. Here as sketched in Fig. \ref{fig3}(a), we introduce external sources (yellow spots) and losses (unavoidable in nature) at different sites to obtain stable excited states that are almost identical to TZMs. The dynamics in the time-domain through external excitation can be captured by 
\begin{equation}
\frac{\partial}{\partial t}\left|\phi\right\rangle=-i(H_{\left|\phi\right\rangle}+H_0)\left|\phi\right\rangle+A\left|S\right\rangle e^{-i\widetilde{\omega}t},\label{eq8}
\end{equation}
where $H_{0}=\sum(-il_{a}|a_{i}\rangle\langle a_{i}|-il_{b}|b_{i}\rangle\langle b_{i}|)$ with $l_a$ and $l_b$ being the losses at different sites, $|\phi\rangle$ is the state reached, $\widetilde{\omega}$ is the excitation frequency,$\ \left|S\right\rangle\equiv\left(0\cdots,S_i,\cdots\right)^T$ represents the distribution of external sources. Here these excitation sources are all located in the linear chain, and the number of excitation sources can be reduced to one [see Fig. S5]. Given $\widetilde{\omega}$ and $|S \rangle$, Fig. \ref{fig3}(b) shows the intensity $|\varphi|^2$ of the steady state reached at frequency $\widetilde{\omega}$ versus the amplification factor of the sources A. With the increasing of A, Fig. \ref{fig3}(c) shows the waveform of the corresponding excited state, which also undergoes an initial deviation from exponential decay (purple line), emerges an arbitrarily designable plateau (orange and green lines), extends to the entire nonlinear domain (blue line), and eventually diverges at the nonlinear boundary (dark cyan line). When $l_a/v_i\ll1$, the corresponding waveform of the excited states in the nonlinear part can be approximated by\begin{eqnarray}
\nu_ia_i+&\left({\widetilde{\kappa}}_i+\alpha a_{i+1}^2\right)a_{i+1}+\alpha\left(\frac{l_a}{v_i}a_i\right)^2a_{i+1}+l_b\frac{l_a}{v_i}a_i&\approx0,\nonumber\\&Im(a_i)=0,&\label{eq9}
\end{eqnarray}
and all $Re(b_i)=0$. Since the last two terms are pretty small, the excited states and the targeted TZMs are almost identical [See Fig. \ref{fig3}(d) and Fig. S4]. A complementary discussion on the waveform of the excited states in the linear chain (not of interest here) is provided in the Supplemental Material, Sec. 4.

These reconfigurable excited states provide the potential for next-generation nonlinear topological devices. The stability of these excited states against noise in dynamics is critical when considering potential applications. We investigate the stability of $|\varphi\rangle$ by adding a sudden perturbation $|\varepsilon\rangle$ \cite{55Zhou2022}, and then simulating the following evolution of wave function $|\phi\rangle=|\varphi\rangle+|\varepsilon\rangle$. Here $\left|\varepsilon\right\rangle$ is a uniform random noise with values inside the interval $(-3, 3)$ and the one we used is plotted in Fig. \ref{fig3}(e). We implement a pretty large noise amplitude here for demonstration purposes. The effect of the disturbance can be captured by the following similarity function
\begin{eqnarray}
\chi\left(t\right)=\frac{|\langle\phi(t)\left|\varphi\right\rangle|}{\sqrt{\langle\phi(t)\left|\phi\left(t\right)\right\rangle\langle\varphi\left|\varphi\right\rangle}}.\label{eq10}
\end{eqnarray}
In the presence of disturbance,$\chi\left(t\right)$ deviates from 1. Figure \ref{fig3}(f) shows the evolution of $\chi$ for different $A$ in the time domain. In a short time, $\chi$ in all cases returns to $1$. We have also checked different noise configurations, and the results are the same (see Supplemental Material, Sec. 5.). Hence, the excited states are stable in dynamics, which thus exhibits potential for the applications of reconfigurable TZMs.

\section{Conclusions}\label{conclusion}
In summary, we realize arbitrarily configurable TZMs by utilizing nonlinearity. Combining the inherent reconfigurability of nonlinearity and the robustness originating from topology, the nonlinear TZMs are robust against disorders, and can be uniquely controlled by intensity. We show that with a proper excitation scheme, the system can reach a stable steady state that is almost identical to the target TZM. Our model can be readily extended to higher-dimensional systems and implemented within diverse systems, such as circuits \cite{46Hadad2018,49PhysRevLett.123.053902,55Zhou2022}, photonic waveguides \cite{44Mukherjee2020,56Kirsch2021}, mechanical resonators \cite{57PhysRevE.94.022201}, etc.\cite{58PhysRevLett.120.057202,59Pernet2022a}. Thus, we believe our findings will enrich nonlinear topological physics and provide new avenues for compact nonlinear topological devices.

\begin{acknowledgments}
 This work is supported by the National Key Research and Development Program of China (Grant No. 2022YFA1404900), the National Natural Science Foundation of China (Grant No. 12334015, 12274332, 12274330), the China Postdoctoral Science Foundation under Grant Number 2023M74271 and Knowledge Innovation Program of Wuhan-Shuguang (Grant No. 2022010801020125).
\end{acknowledgments}
\bibliography{reference}

\end{document}